\documentstyle [12pt,aaspp]{article}
\def\kms{km ${\rm s}^{-1}$}

\begin{document}

\voffset= 1.05truein
\title{Accurate Wavenumbers for Mid-Infrared\\
Fine-Structure Lines}
\vskip 0.25truein
\author{Douglas M. Kelly\altaffilmark{1}$^,$\altaffilmark{2} and John H.
Lacy\altaffilmark{1}}
\vskip 0.15truein
\affil{Department of Astronomy, University of Texas,}
\affil{Austin, TX 78712}
\affil{email: dkelly@kaya.uwyo.edu}
\altaffiltext{1}{Visiting Astronomer at the Infrared Telescope Facility, which
is operated by the University of Hawaii under contract to the National
Aeronautics and Space Administration.}
\altaffiltext{2}{Current Address:  Department of Physics and Astronomy,
              University of Wyoming, Laramie, WY 82071-3905}

\vskip 1.5truein
\centerline{To be Published in {\sl The Astrophysical Journal Letters}}
\clearpage
\voffset=0.0truein

\begin{abstract}

We present accurate new wavenumbers for a set of 13 mid-infrared
fine-structure lines.  The wavenumbers were determined from
observations of the planetary nebula NGC~7027 and of the red
supergiant $\alpha$~Scorpii.  Most of the new wavenumbers are good
to within 0.0025$\%$, or 8~\kms.  We provide details on the
measurements and present an analysis of the errors.  In addition,
we present the first observations of hyperfine splitting in the
[Na~IV] 1106 cm$^{-1}$ line.

\end{abstract}


\section{Introduction}

The strongest features in the mid-infrared spectra of gaseous nebulae are
the fine-structure lines of [Ne~II], [S~III], [S~IV], and [Ar~III] and
the series of unidentified PAH or carbon grain features.  This set of
atomic and solid state features has been the center of attention for
most mid-infrared galaxy studies to date, and it represents the general
limit of what can be detected in galaxies from the ground with 4-meter class
telescopes (e.g. Roche et al. 1991; Kelly et al. 1995).  However, with
the impending launch of the Infrared Space Observatory (ISO) and with the
development of 8-meter class, infrared-optimized telescopes, new
opportunities are unfolding in the field of mid-infrared spectroscopy.

ISO will be capable of observing a large number of lines in bright
galaxies, including many lines which have never been detected before
except in novae.  In most galaxies, however, the number of detectable
lines will be relatively small so it will be inefficient to get
complete spectral coverage.  It is therefore essential to have
accurate wavenumbers for the lines of interest.  In the course of
this study, we have found that predicted wavenumbers can be in error
by as much as several tenths of a percent.  Thus, in support of the
ISO mission and of our own ground-based programs, we have undertaken
a project to determine accurate new wavenumbers for fine-structure
lines in the mid-infrared.

Our target for most of these observations has been the hot, young
planetary nebula NGC~7027.  This planetary nebula is unusual in that
the central star has a temperature of roughly $2 \times 10^5$ K
(Kaler \& Jacoby 1989), but it is still surrounded by a neutral,
molecular cloud (see Graham et al. 1993).  The kinematics and
morphologies of the various ions in NGC~7027 could be quite
different from each other, and the extended, complicated nature
of NGC~7027 is the major source of our wavenumber uncertainties.

\section{Observations}

We measured a variety of fine structure lines in the mid-infrared
spectrum of NGC~7027 during observing runs at the NASA Infrared
Telescope Facility (IRTF) in 1992 August, 1993 June, and 1994 October.
The observations were made using Irshell, a mid-infrared grating
spectrometer (Lacy et al. 1989; Achtermann 1994).  Irshell has an
11x64 pixel Si:As array, with 11 rows in the spatial direction and
64 columns in the spectral direction.  We used a low dispersion
grating in 1992 August to measure the entire 8-13 $\mu$m spectrum
of NGC~7027 at a resolution of $\lambda/\Delta\lambda\approx1000$.
We switched to an echelle grating to measure individual lines during
the 1993 June and 1994 October runs.  The dispersion of
the echelle observations ranged from 16000-25000 (depending on the
grating angle), and with a 2 pixel slit width (2$^{\prime\prime}$~on
the 3m IRTF), the resulting resolution ranged from 8000-12500.  A
dome-temperature card, chopped against the sky, was used for
flatfielding, atmospheric correction, and fluxing (Lacy et al. 1989).
For the 1994 October run, we installed a test-grade Hughes 20x64
pixel Si:As array.  The 20x64 array has slightly smaller pixels,
and the October 1994 data were measured through a 2 pixel
(1.6$^{\prime\prime}$) wide slit at a resolution of 9500-15000.
We used the bright star $\mu$~Cephei (spectral type M2Ia; Hoffleit
\& Jaschek 1982) for flatfielding, atmospheric correction, and
fluxing of our 1994 October data.

The ideal approach to this project would be to make spectral maps of
each of the lines, and we made such a map for the [S~III] line in 1993
June (see Figure~1).  With these data cubes, we could register the
spatial and spectral positions of the line emission, which would give
us the best chance at determining the rest frame wavenumber of the
emission.  As a bonus, we would also get line intensities and kinematic
information.  However, since time was limited, we had to make a number
of compromises.  First, we observed at only one slit position for each
line.  Since we did not cover the entire source, our information on
line strengths was limited to one-dimensional line intensity profiles.
Second, we did not have the time to be precise in our pointing.  Our
usual practice was to move our N-S slit back and forth a few arcseconds
E-W across the western emission lobe until we found the maximum signal.
Our pointing varied because of guiding and tracking errors and because
the morphologies of the various ions are not identical.  Third, we
rarely measured calibration star spectra.  The card calibration is
adequate in most cases, but a star spectrum is helpful for removing
strong telluric features.  Fourth, we speeded up our observations by
opening the slit to two arcseconds even though we could have improved
our resolution by using a slightly narrower slit.  Fifth, we did not
make long observations.  We spent 2-5 minutes on each of the strong
lines and no more than fifteen minutes on the fainter lines.  In
Section~3, we discuss the effects of these compromises on the accuracy
of our wavenumbers.

The most important result from this paper are the vacuum rest wavenumbers
of the lines.  These values are presented in Table~1, along with
estimated 1$\sigma$~uncertainties (see Section~3)
and relative intensities.  Of particular note are the three hyperfine
components of [Na~IV] at 1106 cm$^{-1}$ (see Section~6).  The reasonable
agreement between the observational wavenumbers and the wavenumbers
predicted by Kaufman \& Sugar (1986) and Wiese et al. (1966) gives us
confidence in our line identifications.  We present a summary of our
wavenumber coverage and non-detections in Table~2.  Spectra for the
lines detected in NGC~7027 are presented in Figure~2. The spectra are
usually set to zero in regions of very poor atmospheric transmission
($<$15 \%).  All data were reduced using the Snoopy data reduction
program (Achtermann 1992).

\section{Uncertainties in Wavenumber Measurements}
\label{sec-uncertainties}

The main sources of error in our wavenumber measurements are summarized
below.  These contributions were summed in quadrature to determine the
uncertainties listed with the wavenumbers in Table~1.  For convenience,
we quote errors in \kms, i.e. as fractions of the wavenumber.

\noindent 1. The wavenumber calibration and when possible the
dispersion of the mid-infrared spectra were determined from the
atmospheric lines.  We determined the centroids of the atmospheric
features from our off-position nod frames, and we were able to
determine the centroids to within 0.5~-~2~\kms.  The errors
introduced by our dispersion solution were always less than
1~\kms.

\noindent 2. We obtained wavenumbers for the atmospheric lines from
the AFGL line list (McClatchey et al. 1973), and the wavenumbers are
good to well under 1~\kms.  However, the atmospheric lines are
often blended, and their relative strengths depend on atmospheric
conditions, so there is some uncertainty in the effective wavenumbers
determined from atmospheric models.  When we had to calibrate our
wavenumber scale from a blend of atmospheric features, our uncertainties
ranged up to 4~\kms.

\noindent 3. Photon noise led to scatter in the measured wavenumbers
of the individual emission lines.  It is almost impossible to
determine the magnitude of this effect because of the velocity
structure of NGC~7027, but we estimate that the errors are negligible
for the strong lines and up to 4~\kms~for the weaker lines.
Since most line measurements involve the averaging of detections at
several positions along the slit, the photon noise is reduced and is
expected to contribute no more than 2~\kms~to the wavenumber
uncertainties.

\noindent 4. The lines in NGC 7027 do not have simple line profiles.
There is often emission at more than one velocity from a given
position in the nebula.  We had limited success at deconvolving the
line profiles, so we instead determined centroids for the lines.  We
were able to measure line centroids that were accurate to
0.5~-~1~\kms.  In some cases, bad flatfielding altered the line
profiles.  In these cases our line centroids could be off by as much
as 5~\kms.  The trio of lines at 1106 cm$^{-1}$ was severely
blended, leading to 3~-~6~\kms~of uncertainty in the line
positions.

\noindent 5.  We hinge our velocity reference frame on the [S~III]
line.  The 3~\kms~uncertainty in the [S~III] wavenumber is
propagated into the error estimate for all of the other lines.
The previous best measurements of the [S~III] wavenumber were by
Baluteau et al. (1976), who found a wavenumber of $534.39 \pm 0.01$
cm$^{-1}$ based on measurements of the Orion Nebula, and by Greenberg,
Dyal, \& Geballe (1977), who found a wavenumber of $534.41 \pm 0.03$
cm$^{-1}$ from observations of three planetary nebulae, including
NGC~7027.  We used three methods to estimate the [S~III] wavenumber.
First, by measuring the mean wavenumber over the [S~III] map and
correcting for the heliocentric velocity of 8.8 $\pm$ 0.6 \kms~given
by Schneider et al. (1983), we determine an average wavenumber of
$534.388 \pm 0.004$ cm$^{-1}$.  Second, by comparing the intensity
and velocity profiles of the [S~III] and [Ne~II] emission, we find
the best match if the [S~III] wavenumber is $534.388 \pm 0.005$
cm$^{-1}$.  The rest wavenumber for [Ne~II] was measured in the
laboratory by Yamada, Kanamori, \& Hirota (1985), and is $780.424
\pm 0.001$ cm$^{-1}$.  Third, by comparing the velocities of the
bright ridges in our [S~III] map to those in the H76$\alpha$
velocity map by Roelfsema et al. (1991), we determine a wavenumber
of $534.387 \pm 0.005$ cm$^{-1}$.  These three values combine to
form our new [S~III] wavenumber estimate of $534.387 \pm 0.005$
The two [S~III] slit positions that pass through the center of the
nebula have flux-weighted average velocity shifts of roughly -1~\kms~
relative to the nebular mean.

\noindent 6. Telescope tracking errors and pointing uncertainties
could have a strong influence on our line profiles and measured
wavenumbers.  We see velocity gradients of as large as 6~\kms~per
arcsecond along the slit, with a total shift of 15~-~20~\kms~in
the line centroids from one end of the slit to the other.  In the
E-W direction, we see a shift of over 10~\kms~to the red in the
flux-weighted mean wavenumber along the slit as the slit is moved
from west to east across the nebula, with most of the shift taking
place in the faint outskirts of the nebula.  We determined our N-S
and E-W positions for each line by comparing the intensity and
velocity profiles along the slit to the [S~III] intensity and
velocity map presented in Figure~1.  These comparisons allowed us
to determine the slit positions to within 2$^{\prime\prime}$, which
allowed us to remove the velocity offsets relative to the [S~III]
frame with 2~\kms~uncertainty. On the basis of these comparisons,
we estimate that the velocity uncertainties due to internal motions
in NGC~7027 are 2~-~4~\kms.  In cases where there was not enough
flux to determine the profile of the line emission along the slit,
we estimate our uncertainties to be 4~-~6~\kms.

\section{[Fe~II] Measurements in Alpha Scorpii}

We observed two [Fe~II] lines in the mid-infrared spectrum of the
red supergiant $\alpha$~Scorpii.  These observations will be
discussed by Haas et al.~(1996).  The wavenumbers were reduced to the
rest frame using the heliocentric velocity of -5.0 $\pm$ 0.6~\kms~
determined by Brooke, Lambert, \& Barnes (1974) for low-excitation
metals in the spectrum of $\alpha$~Scorpii.  The wavenumber
uncertainties are dominated by uncertainties in the effective
wavenumbers of blended atmospheric lines.  The [Fe~II] wavenumbers are
listed in Table 1.

\section{Line Intensities}

Our slit measurements provide us with intensity measurements for the
mid-infrared lines in NGC~7027.  As can be seen from the [S~III] map
in Figure~1, the brightest emission for the low ionization lines
comes from the bright ridges to either side of the nebula center (see
also Aitken \& Roche 1983).   Our slit profiles for the higher
ionization lines are consistent with their having similar spatial
distributions.  Since most of our line measurements were made with
our N-S slit crossing one or both lobes, we have reasonable
estimates for the peak intensities of the lines.  These peak
intensities are listed in Table~1.  It should be emphasized that
these intensities do not all refer to identical spatial positions,
and there is a factor of two variation in the continuum levels for
the various lines.  No corrections have been made for extinction.
These intensities are provided only as a general guideline to the
strengths of mid-infrared lines in photoionized gas.  Line ratios
using these numbers could easily be off by a factor of two.

\section{Hyperfine Splitting of the [Na~IV] Line}
\label{sec-hyperfine}

Since the dominant isotope of sodium, $^{23}$Na, has nuclear spin,
we suspect that the group of lines at 1106 cm$^{-1}$ are the
hyperfine components of the [Na~IV] $^3P_2~-~^3P_1$ transition, which
is predicted to lie near this wavenumber.  To test this hypothesis,
we calculated the expected hyperfine splitting and relative line
intensities.  The hyperfine energy splittings are due mainly to the
interaction between the magnetic dipole moment of the nucleus and
the magnetic field at the nucleus due to the orbital and spin motion
of the electrons.  There is a secondary contribution due to the
interaction between the quadrupole electric field of the nucleus and
the electrons.  Hyperfine splitting has not previously been observed
in [$^{23}$Na~IV] but was observed by Harvey (1965) in the isoelectronic
atom [$^{17}$O~I].  To calculate the [Na~IV] hyperfine splittings, we
used the fact that both fine-structure and hyperfine structure
splittings are proportional to $<r^{-3}>$ to scale from the hyperfine
splitting measurements of [$^{17}$O~I] and the fine-structure splittings
of [O~I] and [Na~IV], taking into account the differing nuclear magnetic
dipole and electric quadrupole moments and the fact that the fine-structure
splittings are also proportional to Z$_i$.  We calculated the expected
line intensities using the formulae presented in Townes \& Schawlow
(1955).  There are eight hyperfine components in four closely spaced
groups, from the four hyperfine sublevels of the J=2 fine-structure
level.  Of these, the two highest wavenumber groups are blended
at our resolution, leaving three resolved lines.  The resulting
spectrum (shown as a dotted curve in the [Na~IV] spectrum of Figure 2)
agrees remarkably well with the data, confirming our identification of
this multiplet of lines.

We note that $^{25}$Mg, $^{39}$K, and $^{41}$K also have nuclear spin,
but the observed $^3P_0~-~^3P_1$ transitions should have unresolvably
small hyperfine splittings.  However, the hyperfine splitting of the
strong $^3P_2~-~^3P_1$ (5.6$\mu$m) line of [$^{25}$Mg~V] should be
resolvable, allowing a determination of the $^{25}$Mg/$^{24}$Mg
abundance ratio.  Several isotopic K and Cl fine-structure lines
may also have resolvable hyperfine structure.

\acknowledgements
We are very grateful to Matt Richter and Kevin Luhman for their
assistance with the observations and to the daycrew at the IRTF for
their excellent instrument support.  We thank M. Haas, M. Werner,
and E. Becklin for allowing us to present the [Fe~II] wavenumbers.
Special thanks to Jeff Achtermann for developing the Snoopy data
reduction software and the DSP-based Linus operating system.  This
work has been supported by USAF contract F19628-93-K-0011 and by
NSF grant AST-9020292.

\clearpage
\begin{table}
\begin{center}
\centerline{\bf TABLE 1}
\centerline{Fine-Structure Line Measurements}
\begin{tabular}{llrlcrr}
\tableline
\tableline
Ion    & Transition      & Obs. Date & ~~~Wavenumber        & Peak          &
$\chi_{lower}$ & $\chi_{upper}$ \cr
       &                 &           & ~~~~~~~(cm$^{-1})$   & Intensity$^a$ &
(eV)           & (eV)           \cr
\tableline
[Fe II]& $a^4F_{5/2}-a^4F_{7/2}$ & 6-8~Jun 93 &~~407.84~ $\pm$ .02 &   &   7.87
&  16.18 \cr
[Ne V] & $^3P_0 - ^3P_1$ &~6~Jun 93 &~~411.256  $\pm$ .006  & 1.3(-11) &  97.11
& 126.21 \cr
[S III]& $^3P_1 - ^3P_2$ &~4~Jun 93 &~~534.387  $\pm$ .005  & 2.6(-12) &  23.33
&  34.83 \cr
[Fe II]& $a^4F_{7/2}-a^4F_{9/2}$ & 6-8~Jun 93 &~~557.50~ $\pm$ .02 &   &   7.87
&  16.18 \cr
[Mg V] & $^3P_1 - ^3P_0$ &~9~Jun 93 &~~739.581  $\pm$ .012  & 1.5(-12) & 109.24
& 141.27 \cr
[Ar V] & $^3P_0 - ^3P_1$ &20~Aug 92 &~~763.23~  $\pm$ .05   & 1.3(-12) &  59.81
&  75.02 \cr
[Ne II]& $^2P_{3/2}-^2P_{1/2}$&~4~Jun 93 &~$^b$780.4238$\pm$.001&2.5(-12)&21.56
&  40.96 \cr
[Na IV]& $^3P_{2(F=7/2)}-^3P_1$&10~Jun 93 &1105.88~ $\pm$ .02 &3~~(-13)&  71.64
&  98.91 \cr
       & $^3P_{2(F=5/2)}-^3P_1$&10~Jun 93 &1106.12~ $\pm$ .02 &2~~(-13)&  71.64
&  98.91 \cr
       & $^3P_{2(F=3/2)}-^3P_1$&10~Jun 93 &1106.30~ $\pm$ .03 &1~~(-13)&  71.64
&  98.91 \cr
       & $^3P_{2(F=1/2)}-^3P_1$&10~Jun 93 &                 &          &  71.64
&  98.91 \cr
[Mg VII]&$^3P_0 - ^3P_1$ &10~Jun 93 & 1110.00~  $\pm$ .05   & 8~~(-14) & 186.51
& 224.95 \cr
[Ar III]&$^3P_2 - ^3P_1$ &10~Jun 93 & 1112.176  $\pm$ .015  & 5.1(-12) &  27.63
&  40.74 \cr
[K VI] & $^3P_0 - ^3P_1$ &10~Jun 93 & 1132.52~  $\pm$ .03   & 1.0(-13) &  82.66
&  99.89 \cr
[Ar V] & $^3P_1 - ^3P_2$ &24~Oct 94 & 1265.57~  $\pm$ .04   & 1.9(-12) &  59.81
&  75.02 \\
\tableline
\multicolumn{7}{l}{$^a$erg~s$^{-1}$~cm$^{-2}$~sr$^{-1}$; not corrected for
extinction; uncertainties can be large (see Section 5)}\\
\multicolumn{7}{l}{$^b$wavenumber determined from laboratory measurements by
Yamada et al. (1985)}\\
\end{tabular}
\end{center}
\end{table}
\clearpage
\begin{table}
\begin{center}
\centerline{\bf TABLE 2}
\centerline{Non-Detection Summary}
\begin{tabular}{llclrl}
\tableline
\tableline
Target Line &\multicolumn{3}{c}{Wavenumber
Coverage}&\multicolumn{2}{l}{~~Non-Detections} \cr
            &\multicolumn{3}{c}{(cm$^{-1})$}        &          &
\cr
\tableline
[Fe II]     & ~~~~406.25&\makebox[0pt]{-}&  ~408.60 &          &
\cr
[Ne V]      & ~~~~410.75&\makebox[0pt]{-}&  ~413.30 & [Ca VI]  & ~411.5$^a$
\cr
[S III]     & ~~~~533.8 &\makebox[0pt]{-}&  ~535.1  &          &
\cr
[Fe II]     & ~~~~556.75&\makebox[0pt]{-}&  ~558.50 &          &
\cr
[Mg V]      & ~~~~735.22&\makebox[0pt]{-}&  ~741.1  & [Fe III] & ~739.1
\cr
[F V]       & ~~~~742.75&\makebox[0pt]{-}&  ~746.99 & [F V]    & ~746.3
\cr
[Ar V]      & ~~~~759.66&\makebox[0pt]{-}&  ~767.12 & [S V]    & ~767.5$^a$
\cr
[Ne II]     & ~~~~778.9 &\makebox[0pt]{-}&  ~781.6  &          &
\cr
[S III]     & ~~~~831.70&\makebox[0pt]{-}&  ~835.18 & [S III]  & ~832.5$^b$
\cr
[Cl IV]     & ~~~~848.93&\makebox[0pt]{-}&  ~853.94 & [Cl IV]  & ~849.6
\cr
[Al VI]     & ~~~1095.25&\makebox[0pt]{-}&  1099.78 & [Al VI]  & 1097.0
\cr
[Na IV]~$+$ & ~~~1104.7 &\makebox[0pt]{-}&  1113.0  &          &
\cr
[Mg VII]~$+$&           &                &          &          &
\cr
[Ar III]    &           &                &          &          &
\cr
[K VI]      & ~~~1129.9 &\makebox[0pt]{-}&  1133.4  &          &
\cr
[Na VI]     & ~~~1159.33&\makebox[0pt]{-}&  1165.20 & [Na VI]  & 1161.4
\cr
            &           &                &          & [Mg IX]  & 1162.3$^a$
\cr
            &           &                &          & [Cl VI]  & 1165.4$^a$
\cr
[Ar V]      & ~~~1263.4 &\makebox[0pt]{-}&  1266.0  &          &
\cr
            & ~~~1268.07&\makebox[0pt]{-}&  1270.81 &          &
\cr
\tableline
\multicolumn{6}{l}{$^a$line is from an excited term and so is expected to be
very weak}\cr
\multicolumn{6}{l}{$^b$electric quadrupole transition is expected to be very
weak}\\
\end{tabular}
\end{center}
\end{table}

%
\clearpage

\noindent{\bf Figure 1:} Intensity map for the continuum-subtracted
[S~III] emission in NGC 7027.  The contour interval is
0.172~erg~s$^{-1}$~cm$^{-2}$~(cm$^{-1}$)$^{-1}$~sr$^{-1}$.  The two contour
lines in the center of the nebula denote a low flux region.  North is
roughly 6 degrees west of vertical in this plot.  Velocities are
indicated for each of the points.  The velocity uncertainties are
4~\kms.

\noindent{\bf Figure 2:} Dereddened spectra for the twelve emission
lines detected in NGC~7027.  Wavenumbers are presented in Table~1
and uncertainties are discussed in Section~3.
The dotted line in the [Na~IV] plot denotes our model fit for the
hyperfine splitting.  The [Na~IV] energy splittings were determined
by scaling from the hyperfine measurements of [$^{17}$O~I] by Harvey
(1965) and from the fine-structure splittings of [O~I] and [Na~IV].
Relative line intensities were calculated using the formulae of
Townes \& Schawlow (1955).  The spectrum has been convolved to have
the same resolution as the [Na~IV] data.

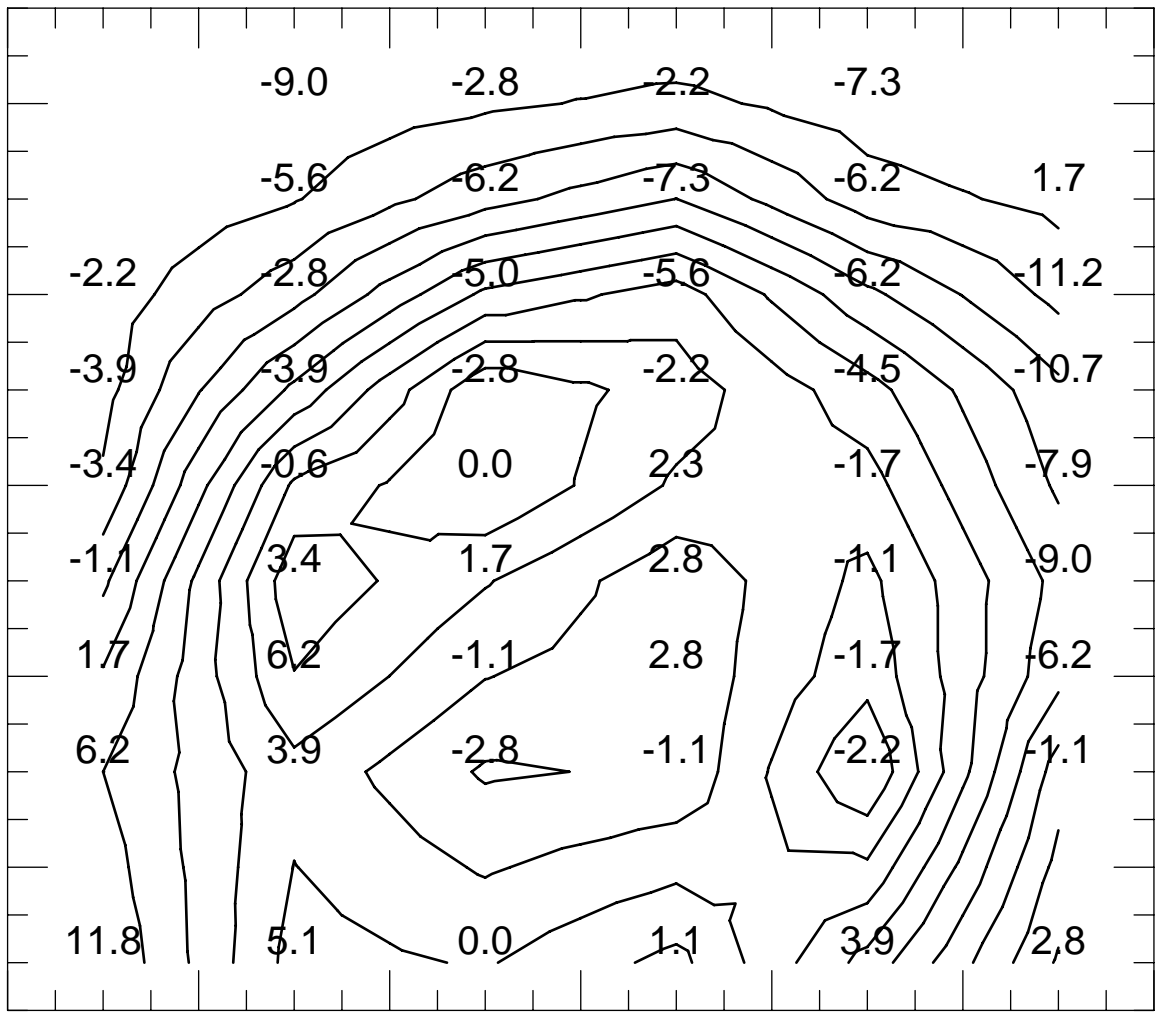
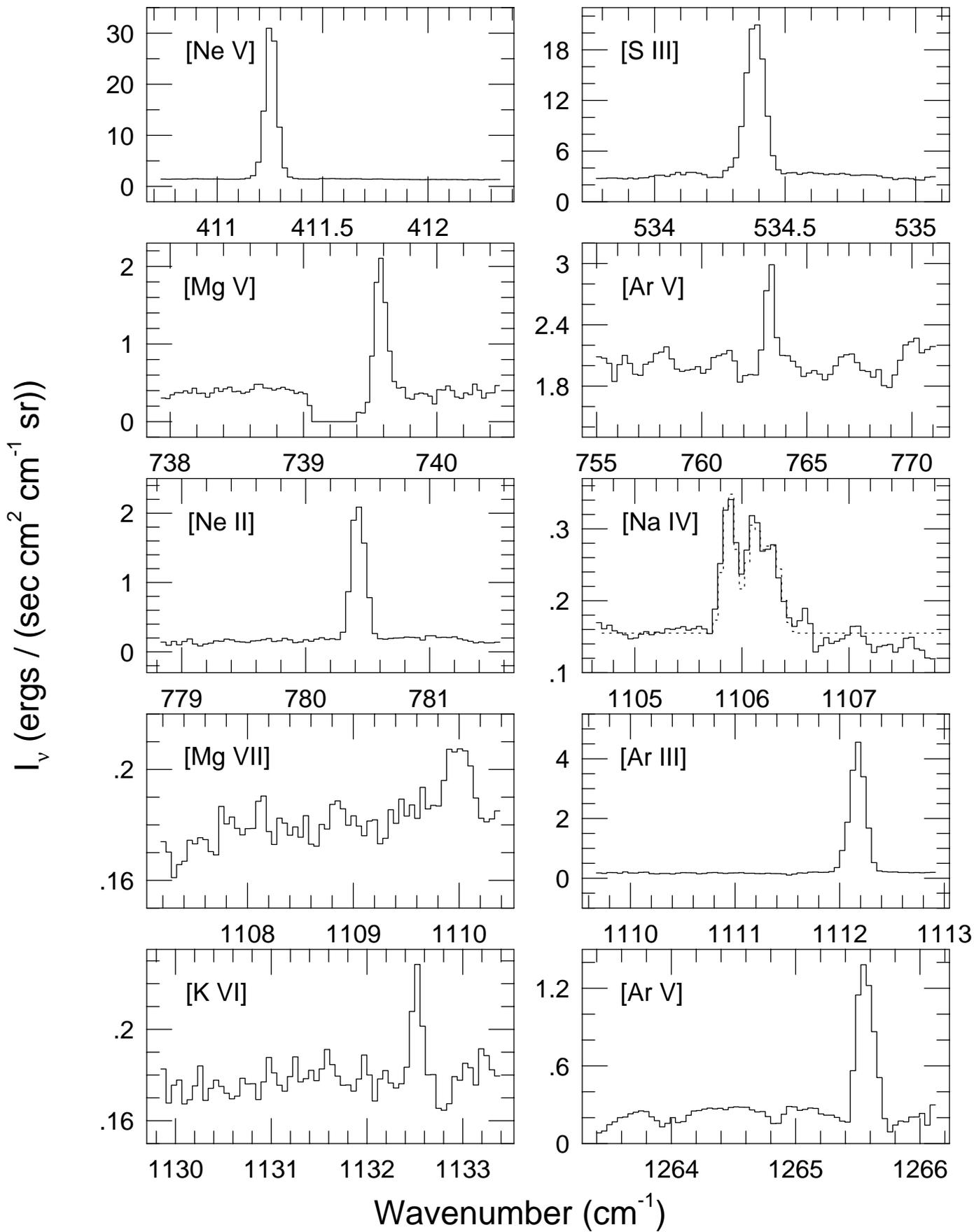
\end{document}